\documentclass[aps,preprint,amsmath,amssymb]{revtex4}
\usepackage{graphicx}
\begin{document}

\title{Search for the Anomalous Interactions of Up-Type Heavy
Quarks in $\gamma\gamma$ collision at the LHC}

\author{M. K\"{o}ksal}
\email[]{mkoksal@cumhuriyet.edu.tr} \affiliation{Department of
Physics, Cumhuriyet University, 58140, Sivas, Turkey}

\author{S.C. \.{I}nan}
\email[]{sceminan@cumhuriyet.edu.tr} \affiliation{Department of
Physics, Cumhuriyet University, 58140, Sivas, Turkey}

\begin{abstract}
We investigate the anomalous interactions of heavy up-type quark $t'$
 in a $\gamma\gamma$ collision at the LHC. We have obtained
$95\%$ confidence level (C.L.) limit of $t'q\gamma$ $(q=u,c)$ anomalous
coupling by taking into account three forward detector acceptances;
$0.0015<\xi<0.15,
 0.0015<\xi<0.5$ and $0.1<\xi<0.5 $.
\end{abstract}


\maketitle

\section{Introduction}
The Standard Model (SM) ensures a conspicuously successful
description of high energy physics at an energy scale of up to a
few hundred GeV. However, the number of fermion families is
arbitrary in the SM. The only limitation on
number of fermion families comes from asymptotic freedom $N\leq8$. We should use at least three fermion families to
obtain CP violation \cite{1} in the SM. CP violation could explain
the matter-antimatter asymmetry in the universe. The SM
with three families is not enough to show the reel magnitude
for matter-antimatter asymmetry of universe. However, this
problem can be solved when the number of family reaches
four \cite{2}. Also, the existence of three or four families is equally
consistent with the updated electroweak precision data \cite{3,4}.
The possible discovery of the fourth SM family may help to
respond to some unanswered questions about electroweak
symmetry breaking \cite{5,7}, fermion's mass and mixing pattern \cite{8,9,10}, and flavor structure of the SM \cite{11,12,13,14}.

Higgs boson is a theoretical particle that is suggested by
the SM. Many experiments were conducted so far to detect
Higgs boson. A boson consistent with this boson was a
detected in $2012$, but it may take quite time to demonstrate
certainly whether this particle is indeed a Higgs boson. If
the lately surveyed $125$ GeV boson is Higgs boson of the
SM \cite{15,16}, the presence of the fourth family would be
disfavoured \cite{17,19}. Besides, a theory with extended Higgs
sector beyond the SM \cite{20} can still include a fourth fermion family even though the $125$ GeV boson is one of the forecasted
extended Higgs bosons. Moreover, the other models estimate
the presence of a heavy quark as a partner to the top quark
\cite{21,22}. Current bounds on the masses of the fourth
SM fermion families are given follows; $m_{t'}>670$ GeV \cite{23},
$m_{b'}>611$ GeV \cite{24}, $m_{l'}>100.8$ GeV, $m_{\nu'}>90.3
(80.5)$ GeV for Dirac (Majorana) neutrinos \cite{25}. When we analyze our
results we have taken into account LHC limits in $\sqrt{s}=7$ TeV. For this purpose, we have assumed $t'$
mass to be greater than
its current experimental limits. The fourth SM quarks would
be produced abundantly in pairs at the LHC via the strong
interaction for masses below O($1$ TeV) \cite{26,27,28,29}, with fairly
large cross sections. The exact designation of their properties
can ensure important advantage in the determination of
new physics which is established upon high energy scales.
Moreover, we can expect a crucial addition from anomalous
interactions for production of fourth family quarks. These interactions have been investigated at lepton colliders
at lepton colliders \cite{30,31}, $\gamma$e colliders \cite{32}, ep colliders \cite{10,33,34,35}, and hadron colliders
\cite{8,28,36,37,38,39,40,41,42,43,44,45}.

The LHC has high energetic proton-proton collisions
with high luminosity. It provides high statistics data. We
expect that this collider will answer many open questions in
particle physics. Research of exclusive production of proton-proton
interactions opens a new field of surveying high
energy photon-induced reactions such as photon-photon and
photon-proton interactions. ATLAS and CMS Collaborations established a program of forward physics with new detectors
located in a region almost $100$ m- $400$ m from the central
detectors. These detectors are called very forward detectors.
They can detect intact protons which are scattered after the
collisions. Very forward detectors can label intact protons
with some momentum fraction loss given the formula $\xi=(|\overrightarrow{p}|-|\overrightarrow{p}'|)/|\overrightarrow{p}|$.
Here $\overrightarrow{p}'$ is the momentum of intact scattered
proton and $\overrightarrow{p}$ is the momentum of incoming proton.
ATLAS Forward Physics Collaboration (AFP)
proposed an acceptance of $0.0015<\xi<0.15$ for the
forward detectors \cite{46}. Two types of measurements will be
planned to examine with high precision using the AFP \cite{47,48,49}: first, exploratory physics (anomalous couplings between
$\gamma$ and $W$ or $Z$ bosons, exclusive production, etc.) and second,
standard QCD physics (double Pomeron exchange, exclusive
production in the jet channel, single diffraction, $\gamma\gamma$ physics,
etc.). These studies will develop the HERA and Tevatron
measurements to the LHC kinematical region. Also, CMS-TOTEM
forward detector scenario has acceptance regions $0.1<\xi<0.5$ and
$0.0015<\xi<0.5$ \cite{50,51}. The TOTEM
experiment at the LHC is concentrated on the studies of the
total proton-proton cross-section, the elastic $pp$ scattering,
and all classes of diffractive phenomena. Detectors housed in
Roman Potswhich can bemoved close to the outgoing proton
beams allow to trigger on elastic and diffractive protons
and to determine their parameters like the momentum loss
and the transverse momentum transfer. Moreover, charged
particle detectors in the forward domains can detect nearly all
inelastic events. Together with the CMS detector, a large solid
angle is covered enabling precise studies \cite{52,53,54}.The forward
detectors of ATLAS and CMS were not built in the first
phase of the LHC. However, the CMS forward detectors were
commissioned in $2009$. The first measurement of the forward
energy flow has been carried out and forward jets at $|\eta|>3$
have been analyzed for the first time at Hadron Colliders
\cite{55}. Also, two photon reactions $pp\rightarrow p\gamma \gamma p\rightarrow p \mu^{-} \mu^{+}p$,
$pp\rightarrow p\gamma \gamma p\rightarrow p e^{-} e^{+}p$ were examined with the help of
forward detectors by the CMS Collaboration in 2012 [56, 57].
On the other hand, AFP Collaboration has not yet installed
the forward detectors. The forward detectors are planned
to be built $210$ m away from the central detectors in $2013$.
Additionally, $420$ m additional detectors will be installed if
physics motivates it later \cite{58}. Forward detectors allow to
determine high energy photon-photon process. This process
occurred by two almost real photons with low virtuality
emitted from protons. The proton structure does not spoil
in this process due to low virtuality of photons. Therefore,
intact scattered protons after the collision can be detected by
the aid of the forward detectors. Searching new physics via
photon-induced reactions have been studied in earlier works
\cite{59,60,61,62,63,64,65,66,67,68,69}.

Photon-photon interaction can be explained by equivalent photon
approximation \cite{70,71}. Emitted photons by protons are
produced an $X$ object via  $pp \rightarrow p \gamma\gamma p
\rightarrow pXp$  process. The cross section of this
process can be found by

\begin{eqnarray}
d\sigma=\int\frac{dL^{\gamma \gamma}}{dW} d \hat{\sigma}_{\gamma
\gamma\rightarrow X} (W) dW
\end{eqnarray}
where $W$ is the invariant mass of the two photon system,
$\hat{\sigma}_{\gamma \gamma\rightarrow X}$ is the cross section for
subprocess $\gamma \gamma\rightarrow X$ and $\frac{dL^{\gamma
\gamma}}{dW}$ is the luminosity spectrum of photon-photon
collisions. $\frac{dL^{\gamma
\gamma}}{dW}$ can be given as follows [63]:

\begin{eqnarray}
\frac{dL^{\gamma
\gamma}}{dW}=\int^{Q_{max}^{2}}_{Q_{1,min}^{2}}dQ_{1}^{2}\int^{Q_{max}^{2}}_{Q_{2,min}^{2}}dQ_{2}^{2}\int^{y_{max}}_{y_{min}}dy\frac{W}{2y}f_{1}\left(\frac{W^{2}}{4y},Q_{1}^{2}\right)f_{2}\left(y,Q_{2}^{2}\right)
\end{eqnarray}
with

\begin{eqnarray}
y_{min}=MAX(W^{2}/(4\xi_{max}E,\xi_{min}E)), y_{max}=\xi_{max}E,
Q_{max}^{2}=2 GeV^{2}
\end{eqnarray}
here $f_{1}$ and $f_{2}$ are functions of equivalent photon
energy spectrum. The photon spectrum with energy $E_{\gamma}$ and virtuality
$Q^{2}$ is given by the following [70]:

\begin{eqnarray}
f=\frac{dN}{dE_{\gamma}dQ^{2}}=\frac{\alpha}{\pi}\frac{1}{E_{\gamma}Q^{2}}\left[(1-\frac{E_{\gamma}}{E})(1-\frac{Q^{2}_{min}}{Q^{2}})F_{E}+\frac{E_{\gamma}^{2}}{2E^{2}}F_{M}\right]
\end{eqnarray}
where

\begin{eqnarray}
Q_{min}^{2}=\frac{m_{p}^{2}E_{\gamma}^{2}}{E(E-E_{\gamma})},
F_{E}=\frac{4m_{p}^{2}G_{E}^{2}+Q^{2}G_{M}^{2}}{4m_{p}^{2}+Q^{2}},
\nonumber\\
G_{E}^{2}=\frac{G_{M}^{2}}{\mu_{p}^{2}}=\left(1+\frac{Q^{2}}{Q^{2}_{0}}\right)^{-4},
F_{M}=G_{M}^{2},    Q^{2}_{0}=0.71 GeV^{2}.
\end{eqnarray}
The terms in above equations are the following: $E$ is the energy of the proton beam which is related to the photon
energy by $E_{\gamma}=\xi E$, $m_{p}$ is the mass of the
proton, $F_{M}$ is function of the
magnetic form factor, $F_{E}$ is function of the electric form
factor and $\mu_{p}^{2}=7.78$ is the magnetic moment of the proton.

In this study, we have examined the anomalous interaction
of up-type $t'$ quark via the $p p \rightarrow p \gamma
\gamma p \rightarrow p q \overline{q} p$ ($q=u,c$) process by
considering three forward detector acceptances; $0.0015<\xi<0.15,
 0.0015<\xi<0.5$ and $0.1<\xi<0.5 $.

\section{Anomalous Interaction of $t'$ quark}

The fourth family $t'$ quark can interact with the ordinary quarks $q_{i}$
via SM gauge bosons ($\gamma$, $g$, $Z^{0}$, $W^{\pm}$) . The
lagrangian of this interaction is expressed by

\begin{eqnarray}
L=-g_{e} Q_{t'}\bar{t}'\gamma^{\mu}t'A_{\mu}-g_{s}\bar{t}'T^{a}\gamma^{\mu}t'G_{\mu}^{a} \nonumber\\
-\frac{g_{Z}}{2}\bar{t}'\gamma^{\mu}(g_{V}-g_{A}\gamma^{5})t'Z_{\mu}^{0} \nonumber\\
-\frac{g_{e}}{2\sqrt{2}\sin
\theta_{W}}V_{t'q_{i}}\bar{t}'\gamma^{\mu}(1-\gamma^{5})q_{i}W_{\mu}^{\pm}+h.c.
\end{eqnarray}
where $g_{e}$ is the electromagnetic coupling constant, $g_{s}$ is
the strong coupling constant, $g_{Z}$ is  the weak neutral current
coupling constant, $g_{A}$ and $g_{V}$ are the vector and
axial-vector type couplings of the neutral weak current with $t'$
quark, $T_{a}$ are the Gell-Mann matrices, $Q_{t'}$ is the electric
charge of fourth family $t'$ quark. The vector fields $A_{\mu}$,
$G_{\mu}$, $Z^{0}_{\mu}$ and $W_{\mu}^{\pm}$ represent photon,
gluon, $Z^{0}$-boson and $W^{\pm}$-boson, respectively. Finally, the
$V_{t'Q_{i}}(Q_{i}=d,b,s,b')$ are the elements of the extended
CKM mixing matrix. In [19] they found that the maximum
value of the fourth generation quark mass is $\sim 300$ GeV
for a Higgs boson mass of $\sim 125$ GeV, which is already in
conflict with bounds from direct searches. Therefore, we
have considered that $t'$ is a heavy quark instead of fourth
generation quark. The $t'$ quark is heavier than the top quark.
It is accepted as the heaviest particle, and it is couple the flavor
changing neutral currents, leading to an enhancement in the
resonance processes at the LHC. The interaction Lagrangian for the anomalous
interactions between the fourth family $t'$ quark, ordinary quarks
$u,c,t$ and the gauge bosons $\gamma,g,Z$ is given as follows:

\begin{eqnarray}
L=\sum_{q_{i}=u,c,t}\frac{\kappa_{\gamma}^{q_{i}}}{\Lambda}
Q_{q_{i}}g_{e}\bar{t}'\sigma_{\mu\nu}q_{i}F^{\mu\nu}+ \nonumber\\
\sum_{q_{i}=u,c,t}\frac{\kappa_{Z}^{q_{i}}}{\Lambda}
\frac{g_{Z}}{2}\bar{t}'\sigma_{\mu\nu}q_{i}Z^{\mu\nu}+ \nonumber\\
\sum_{q_{i}=u,c,t}\frac{\kappa_{g}^{q_{i}}}{\Lambda}
g_{s}\bar{t}'\sigma_{\mu\nu}T_{a}q_{i}G^{\mu\nu}_{a}+h.c.
\end{eqnarray}
where $\kappa_{\gamma}$, $\kappa_{Z}$ and $\kappa_{g}$ are the
anomalous couplings with photon, $Z$ boson and gluon, respectively.
$\Lambda$ is a new physics cutoff and
$\sigma_{\mu\nu}=i[\gamma^{\mu},\gamma^{\nu}]/2$; $F^{\mu\nu}$,
$Z^{\mu\nu}$ and $G^{\mu\nu}_{a}$ are the field stress tensor of the
photon, $Z$ boson and gluons, respectively. Jets that originate from light quarks ($u$, $d$, and $s$) differ
from heavy quarks ($c$ and $b$) in the final state at the LHC.
Therefore, anomalous $\kappa_{\gamma u}$ coupling can be distinguished from
$\kappa_{\gamma c}$ coupling via the process $\gamma \gamma\rightarrow q \bar{q}$, if anomalous
couplings $\kappa_{\gamma u}$ are not equal to $\kappa_{\gamma c}$. It can be understood that
the bound on product $\kappa_{\gamma u} \times \kappa_{\gamma c}$ through the process $\gamma \gamma \rightarrow u \bar{c}$
can be also examined. However, we consider that $\kappa_{\gamma u}$ is equal
to $\kappa_{\gamma c}$ in our paper. For the fourth family leptons $\ell' \ell \gamma$ coupling
was calculated in the literature for the photon-photon fusion
at the LHC[72]. Also, $b'q\gamma$ coupling can be examined through
the process $\gamma \gamma \rightarrow q \bar{q}$ ($q=d,s$). But study of the $b'd\gamma$ and
$b's\gamma$ couplings is difficult for this process since $d$ and $s$ quarks
cannot be distinguished from each other.

Using interaction Lagrangian in (7) anomalous decay
widths of $t'$ quarks can be obtained as follows:

\begin{eqnarray}
\Gamma(t' \rightarrow q \gamma)= \frac{2
\kappa_{\gamma}^{2}}{\Lambda} \alpha_{e} Q_{q_{i}}^{2} m_{t'}^{3}.
\end{eqnarray}
where $m_{t'}$ is the mass of the fourth family $t'$ quark and $\alpha_{e}$ is the electromagnetic coupling constant.

The subprocess $\gamma \gamma \rightarrow q \bar{q}$ consists of $t$ and $u$ channel
tree-level SM diagrams. Additionally, there are two Feynman
diagrams containing $t'$ quark propagators in $t$ and $u$ channels.
The whole polarization summed amplitude square of this
process has been calculated as follows:

\begin{eqnarray}
|M|^{2}=8
g_{e}^{4}Q_{q_{i}}^{4}\left(\frac{t}{u}+\frac{u}{t}\right)-64
g_{e}^{4}Q_{q_{i}}^{4}(\frac{\kappa_{\gamma}}{\Lambda})^{2}\left(\frac{u^{2}}{u-m_{t'}^{2}}+\frac{t^{2}}{t-m_{t'}^{2}}\right) \nonumber\\
+128 g_{e}^{4}Q_{q_{i}}^{4} (\frac{\kappa_{\gamma}}{\Lambda})^{4}
\left[\frac{2 s t u m_{t'}^{2}}{(u-m_{t'}^{2})(t-m_{t'}^{2})}+(t
u+m_{t'}^{2}
s)\left(\frac{u^{2}}{(u-m_{t'}^{2})^{2}}+\frac{t^{2}}{(t-m_{t'}^{2})^{2}}\right)\right]
\end{eqnarray}
where $s$, $t$ and $u$ are the Mandelstam variables and we omit the mass
of ordinary quark $(q_{i}=u,c)$.  We have supposed $\sqrt{s}=14$ TeV to be center ofmass energy of the proton-proton
system during calculations.

The leading order background process comes from QCD induced
reactions (pomeron exchange). Pomerons emitted
from incoming protons can interact with each other, and
they can occur at the same final state. However, survival
probability for a pomeron exchange is quite smaller than
survival probability of induced photons.Therefore, pomeron
background is expected to have minor effect on sensitivity
bounds [73, 74].

\begin{figure}
\includegraphics{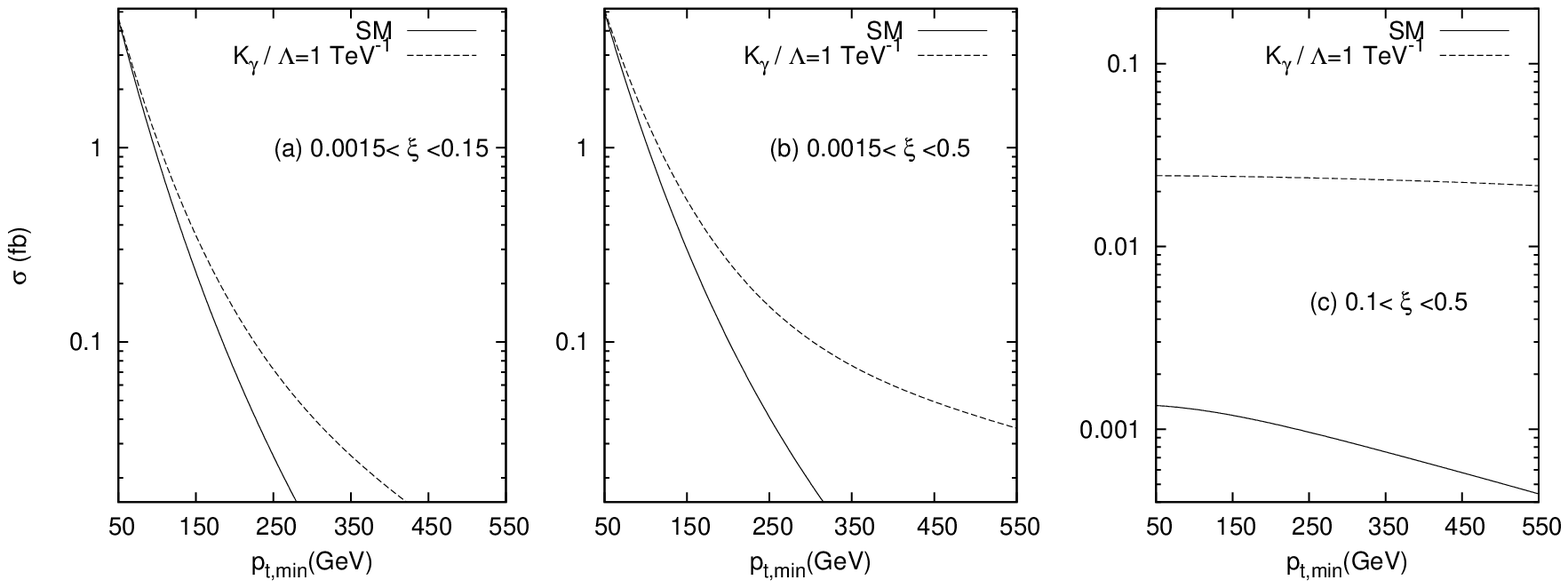}
\caption{The SM and total cross sections of $p p\rightarrow p q
\bar{q} p$ ($q=u,c$) process as a function transverse momentum cut ($p_{t,min}$) on the final state
quarks for three forward detector acceptances: $0.0015<\xi<0.15, 0.0015<\xi<0.5$ and
$0.1<\xi<0.5$. $m_{t'}$ and $\kappa_{\gamma}/{\Lambda}$ is taken to
be $700$ GeV, $1$ TeV$^{-1}$, respectively.
\label{fig1}}
\end{figure}

In Figure (\ref {fig1}), we have plotted the SM and total cross
sections of $p p\rightarrow p q \bar{q} p$ ($q=u,c$) process as a function
 $p_{t,min}$($p_{t}$ cut) transverse momentum of final state quarks for three
forward detector acceptances: $0.0015<\xi<0.15,
 0.0015<\xi<0.5$ and $0.1<\xi<0.5 $. Here $m_{t'}$ and $\kappa_{\gamma}/{\Lambda}$
is taken to be $700$ GeV, $1$ TeV$^{-1}$, respectively. From these
figures, we see that the SM and total cross sections can be
distinguished from each other at large values of the $p_{t}$ cut.
Then, it can be understood that imposing higher values of
$p_{t}$ cut can reduce the SM background. These cuts allow to
obtaining better sensitivity bounds.

In this motivation, we show the SM event numbers of
$p p\rightarrow p q \bar{q} p$ for different values of $p_{t}$ cut and luminosities in Tables 1, 2, and 3 for acceptance regions $0.0015<\xi<0.15,
 0.0015<\xi<0.5$ and $0.1<\xi<0.5 $, respectively. During
statistical analysis we use two different techniques. In the first
approach we apply cuts on the $p_{t}$ of the final state quarks
to suppress the SM cross section. We make the number of
SM event smaller than $0.5$. Then it is very appropriate to set bounds on the couplings using a Poisson distribution since
the number of SM events with these cuts is small enough.
From our calculations, $p_{t}$ cuts are obtained as $380$ GeV and
$452$ GeV for two acceptance regions $0.0015<\xi<0.15$ and $0.0015<\xi<0.5$ in order to be less than $0.5$ the number of SM
event, respectively. Since the invariant mass of the final state
quarks for $0.1<\xi<0.5$ is greater than $1400$ GeV, the SM cross
section is very small. Hence, it does not need a high $p_{t}$ cut
for $0.1<\xi<0.5$ acceptance region. Moreover, ATLAS and CMS have central detectors with a pseudorapidity $|\eta| < 2.5$
for the tracking system at the LHC. Therefore, for all of the
calculations in this paper, we also apply $|\eta| < 2.5$ cut. The
parameter plane $m_{t'}-\kappa_{\gamma}/{\Lambda}$ is plotted at $95\%$ C.L. using Poisson analyse for the three different acceptances:
$0.0015<\xi<0.15, 0.0015<\xi<0.5$ and $0.1<\xi<0.5$ in Figure $2$. In Figures $2(a)$ and $2(b)$, we use the different $p_{t}$ values for every acceptance span to obtain less than $0.5$ event number of SM: a)$p_{t}=380$ GeV for acceptance span $0.0015<\xi<0.15$, b)$p_{t}=452$ GeV for acceptance span $0.0015<\xi<0.5$ as mentioned above. In Figure 2(c), we
applied a $p_{t}$ cut of $p_{t}=50$ GeV for acceptance span $0.1<\xi<0.5$ for detection
of the final state quarks in central detectors.

Second analyze technique, we have used to one parameter $\chi^{2}$ analyze when the SM event number larger than $10$. The $\chi^{2}$ function is given as follows:

\begin{eqnarray}
\chi^{2}=\left(\frac{\sigma_{SM}-\sigma_{NP}}{\sigma_{SM}\delta}\right)^{2}
\end{eqnarray}
where $\sigma_{SM}$ is the cross section of SM, $\sigma_{NP}$ is the cross section containing new physics effects and $\delta=\frac{1}{\sqrt{N_{SM}}}$ is the statistical error.
In Figure $3$, the parameter plane of $m_{t'}-\kappa_{\gamma}/{\Lambda}$ is plotted at $95\%$ C.L. using
$\chi^{2}$ analyse for the two different acceptances $0.0015<\xi<0.15, 0.0015<\xi<0.5$.  For the
$0.1 < \xi < 0.5$ acceptance region we cannot use $\chi^{2}$ analysis
due to SM event number being smaller than 10 as seen from
Table 3. We have found from Figure 3 that $0.0015<\xi<0.5$ acceptance region provides more restrictive limit than
$0.0015<\xi<0.15$ acceptance region because new physics
effect comes from high energy region.

\section{Conclusions}

Forward detector equipments at the LHC can discern intact
scattered protons after the collision. Hence, we can distinguish
exclusive photon-photon processes with respect to deep
inelastic scattering which damages the proton structure. Since
photon-photon interaction has very clean environment, it
is important to examine new physics for a given detector
acceptance region through photon-induced reactions. Moreover,
this interaction can isolate to $\kappa_{\gamma}$ coupling from the
other gauge boson couplings. In these motivations, we have
researched the anomalous interaction of $t'$ quark
via $pp \rightarrow p \gamma\gamma p \rightarrow pXp$ process at the LHC to investigate anomalous
$t'q\gamma$ coupling. Our results show that the sensitivity of the
anomalous $\kappa_{\gamma}/{\Lambda}=0.85$ TeV$ ^{-1}$ coupling can be reached at
$\sqrt{s}=14$ TeV and $L_{int}=100$ fb$^{-1}$ for the $m_{t'}=650$ GeV, the acceptance span $0.0015<\xi<0.5$. As a result, the exclusive $pp \rightarrow p \gamma\gamma p \rightarrow p q \bar{q} p$ reaction at the LHC offers us an important opportunity
to probe anomalous couplings of $t'$ quark.

\pagebreak

\pagebreak

\begin{figure}
\includegraphics{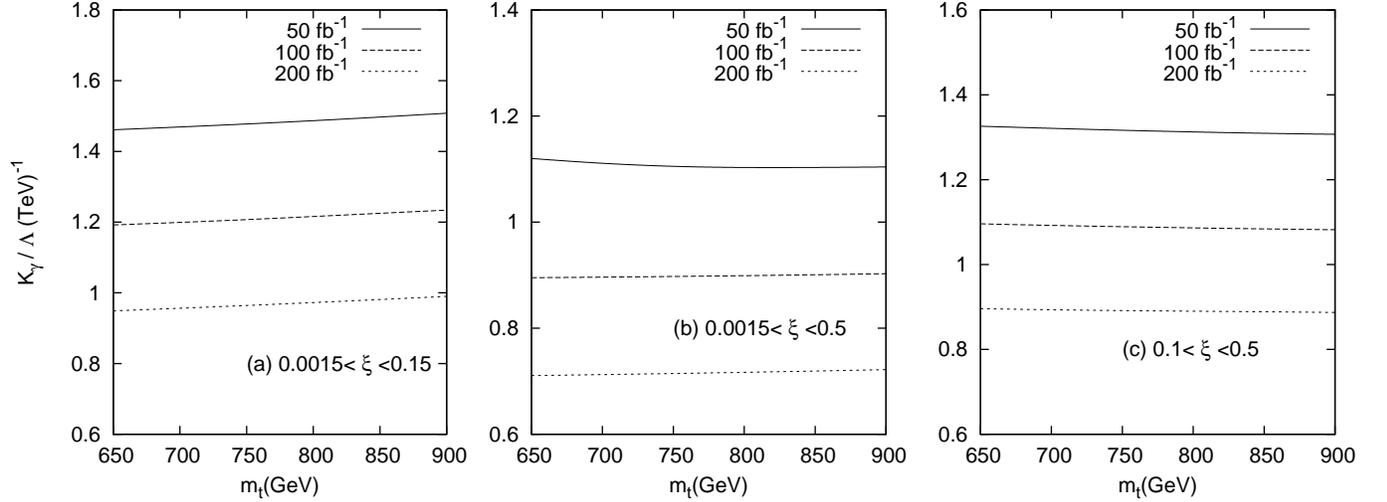}
\caption{The parameter plane of $m_{t'}$ and
$\kappa_{\gamma}/{\Lambda}$ at $95\%$ C.L. using Poisson analysis for three different luminosities:$50$, $100$ and $200$ fb$^{-1}$. In (a)
and (b), we use the different $p_{t}$ values for every acceptance region to obtain less than 0.5 event number of SM: (a) $p_{t}=380$ GeV for acceptance
region $0.0015<\xi<0.15$; (b) $p_{t}=452$ GeV for acceptance region $0.0015<\xi<0.15$. In (c), we applied a $p_{t}$ cut of $p_{t}=50$ GeV for acceptance
region $0.1<\xi<0.5$.
\label{fig2}}
\end{figure}

\begin{figure}
\includegraphics{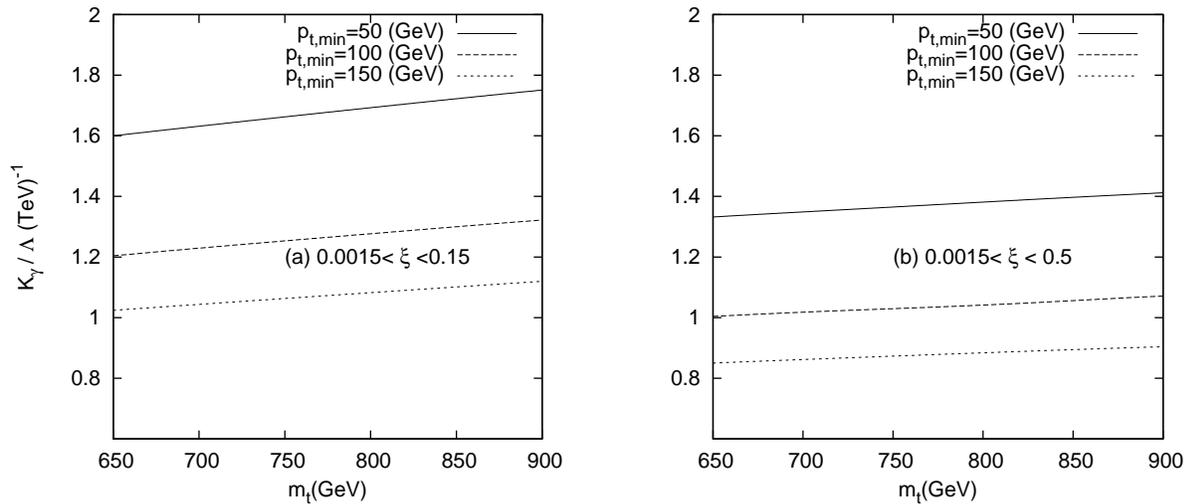}
\caption{The parameter plane of $m_{t'}$ and
$\kappa_{\gamma}/{\Lambda}$ for the two different acceptances: $0.0015<\xi<0.15$ and $0.0015<\xi<0.5$ at $95\%$ C.L. using $\chi^{2}$ analysis. Here, $p_{t,min}$ transverse momentum cuts are taken to be 50, 100, and 150 GeV, respectively\label{fig3}}
\end{figure}

\begin{table}
\caption{The SM event numbers of $p p\rightarrow p q
\bar{q} p$ process for different values of $p_{t}$ transverse momentums and luminosities. Here acceptance span is taken to be $0.0015<\xi<0.15$.
\label{tab1}}
\begin{ruledtabular}
\begin{tabular}{cccc}
$p_{t,min}$(GeV)& 50$fb^{-1}$ &100$fb^{-1}$ &200$fb^{-1}$ \\
\hline
$50$& $206.75$& $413.5$& $827$ \\
$100$& $24.46$& $48.94$& $97.87$  \\
$150$& $5.97$& $11.95$& $23.89$  \\
$200$& $2.02$& $4.05$& $8.11$  \\
$300$& $0.37$& $0.75$& $1.51$  \\
$400$& $0.096$& $0.19$& $0.382$  \\
\end{tabular}
\end{ruledtabular}
\end{table}

\begin{table}
\caption{The SM event numbers of $p p\rightarrow p q
\bar{q} p$ process for different values of $p_{t}$ transverse momentums and luminosities. Here acceptance span is taken to be $0.0015<\xi<0.5$.
\label{tab2}}
\begin{ruledtabular}
\begin{tabular}{cccc}
$p_{t,min}$(GeV)& 50$fb^{-1}$ &100$fb^{-1}$ &200$fb^{-1}$ \\
\hline
$50$& $224.8$& $449.6$& $899.2$ \\
$100$& $29.2$& $58.4$& $116.6$  \\
$150$& $7.8$& $15.6$& $31.3$  \\
$200$& $2.9$& $5.8$& $11.6$  \\
$300$& $0.65$& $1.3$& $2.6$  \\
$400$& $0.21$& $0.42$& $0.83$  \\
$500$& $0.08$& $0.16$& $0.32$  \\
\end{tabular}
\end{ruledtabular}
\end{table}

\begin{table}
\caption{The SM event numbers of $p p\rightarrow p q
\bar{q} p$ process for different values of $p_{t}$ transverse momentums and luminosities. Here acceptance span is taken to be $0.1<\xi<0.5$.
\label{tab2}}
\begin{ruledtabular}
\begin{tabular}{cccc}
$p_{t,min}$(GeV)& 50$fb^{-1}$ &100$fb^{-1}$ &200$fb^{-1}$ \\
\hline
$50$& $0.06$& $0.12$& $0.24$ \\
$100$& $0.057$& $0.115$& $0.23$  \\
$150$& $0.05$& $0.1$& $0.2$  \\
\end{tabular}
\end{ruledtabular}
\end{table}

\end{document}